%% file: paper.tex
\documentclass[sigconf]{acmart}

\usepackage{booktabs}
\usepackage{subcaption}
\usepackage{xspace}
\usepackage{listings}
\usepackage{balance}
\usepackage{enumitem}
\usepackage{tikz}
\usepackage{pgfplots}

\usetikzlibrary{arrows.meta,backgrounds,calc,decorations.pathreplacing,fit,matrix,positioning}
\usepgfplotslibrary{groupplots}
\pgfplotsset{compat=1.18}

\definecolor{ncclblue}{HTML}{2F5D7E}
\definecolor{ncclteal}{HTML}{3D8A86}
\definecolor{ncclgold}{HTML}{C79332}
\definecolor{ncclorange}{HTML}{C46A3C}
\definecolor{ncclred}{HTML}{A84448}
\definecolor{ncclgreen}{HTML}{6E8F4A}
\definecolor{ncclslate}{HTML}{53636F}
\definecolor{ncclink}{HTML}{1F2B37}

\pgfplotsset{
  nccl axis/.style={
    scale only axis,
    axis line style={draw=black!55},
    tick style={draw=black!55},
    tick label style={font=\footnotesize},
    label style={font=\footnotesize},
    title style={font=\footnotesize},
    grid=major,
    grid style={draw=black!10},
    minor tick num=0,
    enlarge x limits=0.04,
    clip=false,
  },
}

\tikzset{
  nccl box/.style={
    rounded corners=3pt,
    draw=black!50,
    very thick,
    fill=white,
    align=center,
    inner sep=5pt,
  },
  nccl soft box/.style={
    rounded corners=4pt,
    draw=black!45,
    very thick,
    fill=black!2,
  },
  nccl chip/.style={
    rounded corners=2pt,
    draw=black!35,
    thick,
    fill=white,
    align=center,
    inner sep=3pt,
  },
  nccl flow/.style={-Latex, line width=1pt, draw=ncclblue},
  nccl feedback/.style={-Latex, line width=1pt, draw=ncclteal},
  nccl alert/.style={-Latex, line width=1pt, draw=ncclorange},
  nccl verdict ok/.style={
    rounded corners=2pt,
    draw=ncclgreen!80!black,
    fill=ncclgreen!18,
    align=center,
    inner sep=3pt,
    minimum width=2.25cm,
  },
  nccl verdict bad/.style={
    rounded corners=2pt,
    draw=ncclorange!90!black,
    fill=ncclorange!20,
    align=center,
    inner sep=3pt,
    minimum width=2.25cm,
  },
}

\newcommand{\sysname}{NCCLbpf\xspace}
\newcommand{\us}{\,\textmu{}s\xspace}
\newcommand{\ns}{\,ns\xspace}

\begin{document}

\title{\sysname: Verified, Composable Policy Execution\\for GPU Collective Communication}

\author{Yusheng Zheng}
\affiliation{\institution{University of California, Santa Cruz}\country{USA}}

\begin{abstract}
NCCL is the de facto standard for collective GPU communication in large-scale distributed training, relying heavily on plugins to customize runtime behavior. However, these plugins execute as unverified native code within NCCL's address space, risking job crashes, silent state corruption, and downtime from restarts during policy updates. Inspired by kernel extensibility models, we introduce \sysname{}, a verified, high-performance extension framework embedding a userspace eBPF runtime directly into NCCL's existing plugin interfaces, without modifying NCCL itself. \sysname{} offers load-time static verification to prevent unsafe plugin execution, structured cross-plugin maps enabling composable policies and closed-loop adaptation, and atomic policy hot-reloads eliminating downtime previously required for policy updates. Evaluations on 8$\times$ NVIDIA B300 GPUs connected via NVLink demonstrate that \sysname{} imposes just 80--130\ns overhead per tuner decision (less than 0.03\% of collective latency), prevents all tested unsafe plugin behaviors at load-time, and enables a message-size-aware eBPF policy that improves AllReduce throughput by up to 27\% over NCCL's default in the 4--128\,MiB range.
\end{abstract}

\maketitle

\section{Introduction}
\label{sec:intro}

Collective GPU communication performance is crucial for large-scale
distributed training frameworks such as
Megatron-LM~\cite{megatron}, DeepSpeed~\cite{deepspeed}, and
PyTorch FSDP~\cite{pytorch-fsdp}, all of which depend on
NCCL~\cite{nccl} for efficient inter-GPU data movement. The
performance of NCCL collectives (e.g., AllReduce, AllGather)
significantly depends on runtime decisions: algorithm selection
(ring vs.\ tree), transport protocols (LL vs.\
Simple~\cite{demystifying-nccl}), and parallelization channels.
Recognizing this dependency, NCCL exposes a modular plugin
architecture, including tuner, profiler, and net plugins, allowing
cloud operators and hardware vendors to tailor policies to match
specific deployment characteristics. Major providers already
leverage these plugins extensively in production
scenarios~\cite{autoccl}.

However, this flexibility comes at substantial reliability and
operability risks. NCCL plugins execute as native code within
NCCL's runtime environment without sandboxing or verification. A
single null-pointer dereference can crash a critical GPU training
job; an infinite loop or race condition can silently halt collective
operations; and subtle memory corruption can cause unpredictable
performance degradation or safety violations. NCCL's own inspector
plugin has caused production training crashes due to use-after-free
and deadlock bugs~\cite{nccl-inspector-bug}, and profiler plugins
have triggered segfaults on multi-GPU
nodes~\cite{nccl-profiler-bug}. Moreover, NCCL plugins lack
structured state-sharing capabilities, making closed-loop policy
adjustments across plugins impossible. Updating plugin logic
requires restarting affected jobs, resulting in disruptive downtime
and hindering rapid policy iteration in production environments.

We identify a structural analogy between NCCL plugin
extension points and kernel-level hooks, where the eBPF
framework~\cite{ebpf-linux} has already successfully addressed
similar challenges~\cite{xrp, electrode, pageflex}. In both
contexts, extension points must execute safely and efficiently in
performance-critical environments. NCCL's net plugin closely
mirrors eBPF's original packet-processing use case, and NCCL's
tuner and profiler interfaces correspond to kernel tracepoints
where lightweight policy-driven decisions influence critical runtime
behavior. Inspired by eBPF's verified extensibility model, we
present \sysname, which integrates a verified, userspace eBPF
runtime~\cite{bpftime} directly within NCCL's existing plugin
interface, without modifying NCCL source code.

\sysname provides three properties: (1)~load-time static
verification guarantees safety before execution, preventing crashes
and state corruption; (2)~structured, typed maps enable composable,
cross-plugin state sharing, enabling previously impossible
closed-loop adaptations; and (3)~atomic policy hot-reload
capability allows operators to update policies seamlessly at
runtime without interruption.

We implement \sysname atop NCCL version 2.29.7 using bpftime, a
lightweight userspace eBPF runtime. Our evaluation on 8$\times$
NVIDIA B300 GPUs with NVLink shows that \sysname adds 80--130\ns
overhead per policy decision, representing less than 0.03\% of
collective latency. Atomic updates incur a downtime of just
1.07\us and introduce zero dropped calls across 400{,}000 policy
invocations. A message-size-aware eBPF policy improves 8-GPU
AllReduce throughput by up to 27\% over NCCL's default algorithm
selection in the 4--128\,MiB range.

This paper makes three contributions:

\begin{enumerate}[leftmargin=*,topsep=2pt,itemsep=1pt]
\item \textbf{\sysname}, the first framework to bring verified,
composable eBPF policy execution to NCCL's plugin architecture,
transparently enhancing safety and extensibility without modifying
NCCL source code
(\S\ref{sec:design}--\ref{sec:impl}).

\item A structured, typed-map-based composability mechanism
enabling previously isolated NCCL plugins to coordinate and
dynamically adapt, facilitating closed-loop policy control
(\S\ref{sec:design}, \S\ref{sec:eval-composability}).

\item An evaluation on 8$\times$ NVIDIA B300 NVLink GPUs
demonstrating low overhead (80--130\ns per decision), verified
runtime safety (rejecting all unsafe plugins tested), and a
message-size-aware policy that improves AllReduce throughput by
up to 27\% (\S\ref{sec:eval}).
\end{enumerate}

\section{Background and Motivation}
\label{sec:motivation}

\paragraph{NCCL and its plugin system.}
NCCL~\cite{nccl} implements GPU collective primitives (AllReduce,
AllGather, Broadcast, etc.) optimized for multi-GPU interconnects.
For each collective call, the \emph{tuner} plugin's
\texttt{getCollInfo} receives the collective type, message size, and
rank topology, and selects an algorithm (ring, tree), protocol (LL,
LL128, Simple~\cite{demystifying-nccl}), and channel count by
modifying a cost table. The \emph{profiler} plugin receives
timestamped event callbacks; the \emph{net} plugin controls network
transport; and the \emph{env} plugin sets runtime parameters. All
four are loaded via \texttt{dlopen} as shared libraries.
The tuner API has evolved rapidly, from v2 (simple integer outputs)
to v5 (in-place 2D float cost table modification) in two
years, and major cloud providers ship custom tuner
plugins~\cite{autoccl}. Despite this growing ecosystem, plugins
remain independent extension points with no cross-plugin
communication and no safety verification.

\paragraph{The safety gap.}
Native plugins execute with full process privileges.
NCCL's own inspector plugin has caused production training crashes
due to use-after-free and deadlock bugs~\cite{nccl-inspector-bug},
and profiler plugins have triggered segfaults on multi-GPU
nodes~\cite{nccl-profiler-bug}. At scale, such failures are costly:
Llama~3 pre-training on 16{,}384~GPUs experienced 466~job
interruptions in 54~days~\cite{llama3}, and analyses of production
GPU clusters find that $\sim$30\% of training jobs fail, with
programming and configuration errors as the dominant
cause~\cite{philly}. The absence of safe hot-reload further
compounds the risk: rolling out plugin updates across a fleet
requires restarting every training job.

\paragraph{eBPF as safe extension mechanism.}
Extended BPF (eBPF)~\cite{ebpf-linux} is an execution framework,
originally implemented in the Linux kernel for packet filtering and
tracing, that statically verifies programs for memory safety and
bounded execution before JIT-compiling them to native code. eBPF has
been adopted as a safe-extension mechanism across
storage~\cite{xrp}, networking~\cite{electrode, etran}, memory
management~\cite{pageflex}, and GPU driver-level resource
management~\cite{gpu-ext}. Recent userspace
runtimes~\cite{bpftime} extend its reach beyond the kernel, enabling
eBPF execution inside application processes without kernel
privileges.
Three properties make eBPF well-suited to NCCL policy execution:
(1)~load-time verification catches memory-safety and termination bugs
that cause the crashes described above; (2)~typed maps provide
concurrent state sharing between profiler and tuner without ad hoc
shared memory; and (3)~atomic program replacement enables verified
hot-reload.

\section{Design}
\label{sec:design}

\subsection{Design Tensions}

\paragraph{T1: Safety vs.\ overhead.}
\sysname verifies only at load time (1--5\,ms, amortized over the job
lifetime), relying on the invariant that verified BPF bytecode, once
JIT-compiled, cannot violate its safety guarantees at runtime. The
tradeoff: verification catches memory safety and termination bugs, not
poor policy decisions.

\paragraph{T2: Structured sharing vs.\ flexibility.}
All cross-component state sharing uses typed eBPF maps with atomic
access semantics. Maps constrain policies to fixed-size keys/values,
but NCCL policy state is inherently simple (per-communicator scalars),
and structured sharing eliminates the locking bugs that plague ad hoc
native plugin state sharing.

\paragraph{T3: Availability vs.\ consistency during updates.}
Hot-reload atomically swaps the function pointer; any in-progress call
completes under the old policy, the next call uses the new one. No
call is lost. If the new policy fails verification, the old policy
continues. The tradeoff: different threads may briefly execute
different policies, which is acceptable because collective decisions
are independent across calls.

\paragraph{T4: Compatibility vs.\ capability.}
Our system operates within NCCL's existing plugin ABI, requiring
no source modifications, no custom builds, and no kernel modules.
The action space is limited to algorithm, protocol, and channel
count selection. Deeper interventions are outside scope, but
the exposed knobs cover the most impactful tuning decisions.
\sysname selects among NCCL's built-in algorithms; it does not
implement custom collective algorithms.

\subsection{Threat Model}
\label{sec:threat-model}

Policies are written by platform operators, not untrusted tenants.
The eBPF verifier (PREVAIL-based) checks memory safety, bounded
execution, stack safety, and helper whitelisting. The system does
\emph{not} protect against resource exhaustion, side channels, bugs in
the trusted computing base (bpftime JIT, PREVAIL verifier, host
plugin), or semantically incorrect decisions. This provides defense in
depth (eliminating crashes, hangs, and memory corruption), analogous
to the Linux kernel's eBPF model.

\subsection{Architecture}

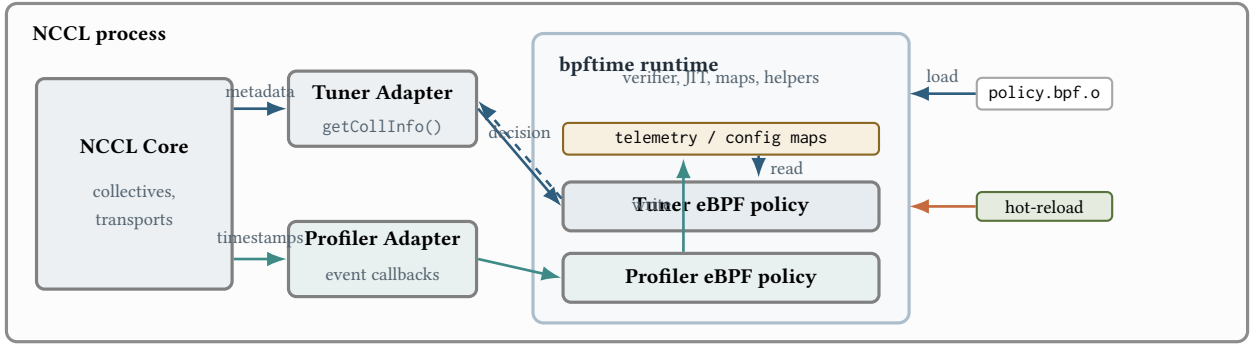
\begin{figure*}[t]
  \centering
  \input{figures/architecture}
  \caption{\sysname architecture. Policy programs are verified via
  PREVAIL, JIT-compiled by bpftime, and execute inside the NCCL
  process. Profiler and tuner communicate through shared typed eBPF
  maps. The entire system loads as standard NCCL plugins.}
  \Description{Architecture diagram showing NCCL Core, Tuner and
  Profiler adapters, bpftime runtime with verifier/JIT/maps/helpers,
  and eBPF policy programs communicating through shared maps.}
  \label{fig:architecture}
\end{figure*}

Figure~\ref{fig:architecture} shows the architecture. Policy authors
write restricted C compiled to BPF ELF objects. At load time, each
program is verified and JIT-compiled to x86-64 code. The map subsystem provides
intra-policy state and cross-plugin communication: a profiler eBPF
program writes latency observations; the tuner reads them. Hot-reload
verifies and JIT-compiles a replacement, then atomically swaps the
function pointer via compare-and-swap.

\paragraph{Policy programming model.}
A policy takes a \texttt{policy\_context} (collective type, message
size, rank count) and writes algorithm, protocol, and channel count
decisions. The verifier ensures policies only read input fields and
write output fields. Map lookups require null checks before
dereference; omitting one causes load-time rejection.

\begin{lstlisting}[language=C, caption={Profiler-to-tuner closed loop.
  The profiler (top) writes latency into a shared eBPF map;
  the tuner (bottom) reads it for adaptive channel selection.
  Null checks (lines 5, 14) are enforced by the verifier.},
  label={lst:policy}, basicstyle=\ttfamily\scriptsize]
/* --- profiler eBPF program --- */
SEC("profiler")
int record_latency(struct profiler_context *ctx) {
  __u32 key = ctx->comm_id;
  struct latency_state *st =
    bpf_map_lookup_elem(&latency_map, &key);
  if (!st) return 0;
  st->avg_latency_ns = ctx->latency_ns;
  st->channels = ctx->n_channels;
  return 0;
}
/* --- tuner eBPF program --- */
SEC("tuner")
int size_aware_adaptive(struct policy_context *ctx) {
  __u32 key = ctx->comm_id;
  struct latency_state *st =
    bpf_map_lookup_elem(&latency_map, &key);
  if (!st) { ctx->n_channels = 4; return 0; }
  if (ctx->msg_size <= 32 * 1024)
    ctx->algorithm = NCCL_ALGO_TREE;
  else
    ctx->algorithm = NCCL_ALGO_RING;
  ctx->protocol = NCCL_PROTO_SIMPLE;
  if (st->avg_latency_ns > 1000000)
    ctx->n_channels = min(st->channels + 1, 16);
  else
    ctx->n_channels = st->channels;
  return 0;
}
\end{lstlisting}

The two programs share \texttt{latency\_map} via the eBPF map
subsystem, a capability absent from NCCL's native plugin
architecture, where tuner and profiler have no shared state.
Null checks on map lookups (lines~5 and~14) are enforced by the
verifier; omitting either causes load-time rejection.

\section{Implementation}
\label{sec:impl}

\sysname is implemented in approximately 2{,}500 lines of C/C++
(plugin host) and 500 lines of restricted C (policy programs).

\paragraph{bpftime integration.}
We embed bpftime~\cite{bpftime} as the userspace eBPF runtime.
bpftime provides an LLVM-based JIT compiler, a
PREVAIL-based~\cite{prevail} static verifier, and a map subsystem.
The verifier runs in the same address space as the plugin, adding
approximately 1--5\,ms of one-time startup overhead amortized over
the lifetime of the training job. The LLVM JIT produces optimized x86-64 code, narrowing the gap
to native performance.

\paragraph{Plugin registration.}
\sysname registers itself as both a tuner and profiler plugin via
NCCL's \texttt{ncclTunerPlugin\_v3} (or v5) and
\texttt{ncclProfilerPlugin\_v1} (or v6) interfaces. The tuner
plugin's \texttt{getCollInfo()} function constructs a
\texttt{policy\_context}, invokes the JIT-compiled eBPF function,
and copies the output fields into NCCL's result structure. The
profiler plugin's event callbacks extract latency information and
update the shared eBPF maps.

\paragraph{NCCL integration challenges.}
NCCL's tuner API uses cost arrays rather than direct algorithm IDs:
the tuner sets costs to zero for preferred choices and to a sentinel
value (\texttt{1e9}) for others, allowing NCCL to fall back
gracefully if the requested combination is unavailable. Our
integration translates eBPF policy outputs into cost array entries.
NCCL also passes a maximum channel count that the tuner must respect;
our native baseline layer clamps the policy's request. Communicator
identification required deriving a stable ID from the context pointer
via hashing, since the communicator ID is not directly exposed in the
tuner API as a simple integer.

\paragraph{Hot-reload mechanism.}
The active policy is stored as an atomic function pointer. Reload has
three phases: (1)~verify the new BPF program, (2)~JIT-compile it,
(3)~atomic compare-and-swap on the pointer. The total reload cost is
$\sim$9.4\,ms; only the final swap (1.07\us) touches the hot path.
The old pointer is retained until in-flight calls drain. If the
replacement fails verification, the swap is aborted and the old policy
continues; the system never enters an unverified state.

\paragraph{Native baseline for comparison.}
To provide a fair overhead comparison, we implemented a native C++
baseline that performs identical policy logic without the eBPF layer,
compiled with the same optimization flags (\texttt{-O2}). The
overhead difference between the native baseline and eBPF policies
directly measures the cost of the eBPF dispatch and JIT execution
layers, isolated from any policy logic cost.

\section{Evaluation}
\label{sec:eval}

\paragraph{Testbed.}
CPU microbenchmarks run on a 240-core AMD EPYC~9575F; each benchmark
performs 1\,million calls and reports P50/P99 latencies. GPU
experiments use 8$\times$ NVIDIA B300 SXM6 GPUs (Blackwell,
275\,GB each) connected via NVLink~5 (NV18, 1.8\,TB/s per GPU)
with CUDA~13.0 and NCCL~2.29.7, hosted on Verda.

\subsection{Overhead}
\label{sec:eval-overhead}

\begin{table}[t]
\centering
\caption{CPU microbenchmark (1M calls each). eBPF policies add
80--130\ns over native code. The overhead decomposes into plugin
framework base (+80\ns, of which eBPF JIT dispatch is 33\ns),
map lookup (+30\ns each), and map update (+10\ns each).}
\label{tab:overhead}
\small
\begin{tabular}{lrrr}
\toprule
\textbf{Policy} & \textbf{P50 (ns)} & \textbf{P99 (ns)} & \textbf{$\Delta$P50} \\
\midrule
Native baseline    &  20 &  30 & --- \\
\texttt{noop}              & 100 & 111 & +80 \\
\texttt{size\_aware\_v2}   & 100 & 111 & +80 \\
\texttt{lookup\_only}      & 130 & 140 & +110 \\
\texttt{lookup\_update}    & 140 & 151 & +120 \\
\texttt{adaptive\_channels}& 140 & 151 & +120 \\
\texttt{slo\_enforcer}     & 150 & 160 & +130 \\
\bottomrule
\end{tabular}
\end{table}

Table~\ref{tab:overhead} reports per-call latency. The overhead
follows an approximate model of $80 + 30n_{\text{lookup}} +
10n_{\text{update}}$\ns (array maps are faster than hash maps,
causing minor deviations). Even the most complex policy
(\texttt{slo\_enforcer}) adds only 130\ns. For a 128\,MiB
8-GPU AllReduce on NVLink (${\sim}$394\us), this is 0.03\% of
collective latency.

End-to-end GPU measurements on NVLink across small message sizes
(8\,B to 256\,KiB) show that the \texttt{noop} plugin adds
${\sim}$1.3\us of fixed overhead (${\sim}$4\% of the ${\sim}$32\us
NVLink baseline). This overhead comes from the plugin framework (shared
memory setup, cost table writes), not the eBPF dispatch itself
(33\ns). For messages of 4\,MiB and above, the overhead is below
measurement noise ($<$0.1\%).

\subsection{Safety and Hot-Reload}
\label{sec:eval-verifier}

We tested 14 eBPF programs against the PREVAIL-based verifier: 7~safe
policies (including all policies in Table~\ref{tab:overhead}) were
accepted; 7~unsafe programs, each targeting a distinct bug class
(null-pointer dereference, out-of-bounds access, illegal helper,
stack overflow, unbounded loop, input-field write, division by
zero), were rejected at load time with actionable error messages.

The safety difference is concrete. We implemented the same
null-dereference bug as both a native C plugin and an eBPF policy:

\begin{lstlisting}[basicstyle=\ttfamily\scriptsize,numbers=none,frame=none,xleftmargin=1em]
Native plugin:  Signal: SIGSEGV (address 0x0)
                  in getCollInfo() at native_bad_plugin.so
eBPF policy:    VERIFIER REJECT: R0 is a pointer to
                  map_value_or_null; must check != NULL
                  before dereference at insn 7
\end{lstlisting}

\noindent The native plugin crashes the training job; the eBPF policy
is caught before execution.

Hot-reload swaps the active policy atomically via compare-and-swap
(1.07\us swap time; $\sim$9.4\,ms including verification and JIT).
Across 400{,}000 continuous invocations, we observe zero lost calls.
If the replacement fails verification, the old policy continues
uninterrupted.

\subsection{Case Studies}
\label{sec:eval-casestudies}

\paragraph{NVLink-aware adaptive policy.}

\begin{figure}[t]
\centering
\begin{tikzpicture}
\begin{axis}[
  nccl axis,
  width=0.92\columnwidth,
  height=5.5cm,
  xlabel={Message Size},
  ylabel={Bus Bandwidth (GB/s)},
  xmode=log, log basis x=2,
  xtick={4,8,16,32,64,128,256,1024,8192},
  xticklabels={4M,8M,16M,32M,64M,128M,256M,1G,8G},
  ymin=0, ymax=900,
  xmin=3, xmax=10000,
  legend style={at={(0.03,0.97)}, anchor=north west,
    font=\small, draw=black!40},
  legend cell align=left,
]
\addplot[color=ncclblue, mark=square*, mark size=2.5pt, thick]
  coordinates {
    (4,132.1) (8,194.9) (16,277.5) (32,348.5)
    (64,425.1) (128,595.3) (256,656.5) (1024,732.7) (8192,836.3)
  };
\addplot[color=ncclgreen, mark=triangle*, mark size=2.5pt, thick]
  coordinates {
    (4,139.8) (8,246.5) (16,336.1) (32,401.6)
    (64,470.8) (128,627.9) (256,656.5) (1024,732.7) (8192,836.3)
  };
\addplot[color=ncclred, mark=x, mark size=3pt, thick, dashed]
  coordinates {
    (4,9.4) (8,17.1) (16,33.7) (32,47.4)
    (64,50.3) (128,38.6) (256,40.1) (1024,43.2) (8192,44.8)
  };
\legend{NCCL default (NVLS), eBPF policy (v2), bad\_channels (1ch)}
\end{axis}
\end{tikzpicture}
\caption{8-GPU AllReduce on NVLink. The eBPF policy selects
Ring/LL128 for 4--32\,MiB and Ring/Simple for 64--192\,MiB,
improving throughput by up to 27\% over NCCL's default NVLS
algorithm. A deliberately bad policy (1~channel) shows that
policies have real control, including destructive power.}
\label{fig:nvlink-policy}
\end{figure}
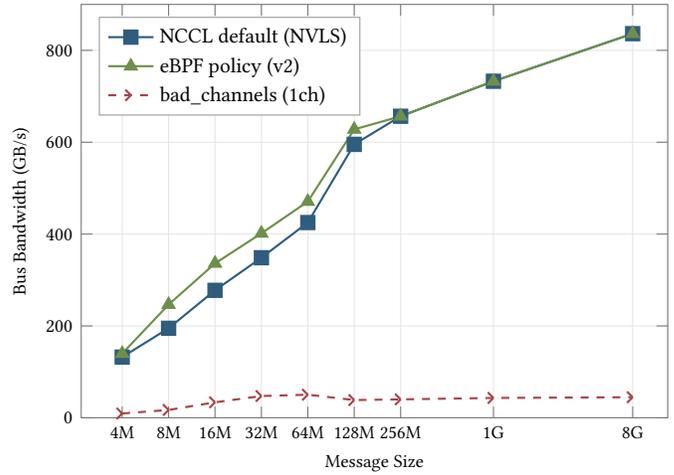

On our 8$\times$~B300 NVLink testbed, NCCL~2.29.7 defaults to the
NVLS algorithm (NVLink SHARP with hardware multicast) for all
message sizes. A parameter sweep reveals that Ring with 32~channels
outperforms NVLS by 5--27\% in the 4--128\,MiB range, while NVLS
is superior at 256\,MiB and above (Table~\ref{tab:sweep}).

\begin{table}[t]
\centering
\caption{Algorithm sweep: 8-GPU AllReduce bus bandwidth (GB/s).
Ring outperforms NVLS by 5--27\% in the 4--128\,MiB range.}
\label{tab:sweep}
\small
\begin{tabular}{lrrr}
\toprule
\textbf{Size} & \textbf{Default (NVLS)} & \textbf{Ring} & \textbf{$\Delta$} \\
\midrule
4\,MiB   & 133.5 & 148.1 & +10.9\% \\
8\,MiB   & 196.3 & 249.7 & +27.2\% \\
16\,MiB  & 278.8 & 337.4 & +21.0\% \\
32\,MiB  & 349.3 & 402.4 & +15.2\% \\
64\,MiB  & 425.2 & 471.8 & +11.0\% \\
128\,MiB & 596.9 & 628.9 & +5.4\% \\
256\,MiB & 656.5 & 632.5 & $-$3.7\% \\
8\,GiB   & 836.3 & 697.6 & $-$16.6\% \\
\bottomrule
\end{tabular}
\end{table}

Static environment variables (\texttt{NCCL\_ALGO=Ring}) cannot
exploit this: they apply globally and would degrade the 256\,MiB+
range. An eBPF policy can select per-message-size. Our
\texttt{nvlink\_ring\_mid\_v2} policy (fewer than 20~lines of C) selects
Ring/LL128 for 4--32\,MiB, Ring/Simple for 64--192\,MiB, and
defers to NCCL's default otherwise.

Figure~\ref{fig:nvlink-policy} shows the result. The eBPF policy
achieves 5.5--26.5\% improvement over NCCL default in the targeted
range (steady-state, after 2--3 warmup communicator creations that
NCCL requires to stabilize Ring/LL128 GPU buffers), while matching
default performance outside this range. The \texttt{noop} policy (not
shown) produces identical throughput to the no-plugin baseline,
confirming zero overhead at 4\,MiB+.

A deliberately bad policy (\texttt{bad\_channels}, forcing 1~channel)
passes the verifier (it is memory-safe) but causes 87--95\%
throughput degradation, demonstrating that policies have real
control over NCCL's behavior. The verifier prevents crashes and
memory corruption, not poor decisions; semantic validation remains
the operator's responsibility.

\paragraph{Stability.}
Over 20~independent runs of 8-GPU AllGather at 128\,MiB, the
default configuration achieves $565.6\pm0.9$\,GB/s
(CV\,=\,0.15\%) and the eBPF v2~policy achieves
$565.5\pm0.6$\,GB/s (CV\,=\,0.10\%). Both configurations are
highly stable on NVLink, with the policy showing 32\% lower
variance. The default has one 3.4$\sigma$ outlier (562.6\,GB/s);
the policy distribution is unimodal with no comparable outlier.

\paragraph{Profiler-to-tuner composability.}
\label{sec:eval-composability}
To evaluate cross-plugin composability, we construct an adaptive
channels policy that starts with a conservative channel count
(nChannels\,=\,2) and relies on profiler telemetry to adapt
upward. Without the profiler, the tuner receives no samples and
remains at 2~channels. With the profiler enabled, latency samples
flow into a shared eBPF map, and the tuner ramps from 2 to
12~channels over 100{,}000 calls. Under injected contention
(10$\times$ latency spike), the policy reduces channels from 12
to~2; upon recovery, it ramps back to~12 within 100{,}000 calls.
This three-phase response (baseline$\to$contention$\to$recovery)
validates \sysname's composability model: two independently
deployed eBPF programs cooperate through shared typed maps, a
capability absent from NCCL's native plugin architecture.

\paragraph{Net plugin extensibility.}
We implement an eBPF-wrapped net plugin that interposes on
NCCL's Socket transport. The wrapper forwards all transport
operations to the built-in backend while executing a BPF program
at each \texttt{isend}/\texttt{irecv} call to count bytes and
connections via a shared eBPF map. The wrapped transport adds
less than 2\% overhead, confirming that eBPF hooks can operate
on NCCL's data-plane path with negligible cost, paralleling eBPF's
established role in kernel packet processing.

\section{Related Work}
\label{sec:related}

\paragraph{Collective communication.}
MSCCL~\cite{msccl} and TACCL~\cite{taccl} synthesize static
schedules offline; they do not address runtime policy selection
or safety. AutoCCL~\cite{autoccl} automates tuning via search
but uses native code without verification.
MSCCL++~\cite{mscclpp} provides GPU-driven primitives at the
mechanism layer. Hu et al.~\cite{demystifying-nccl} provide a
comprehensive analysis of NCCL's protocol variants and algorithms.
gpu\_ext~\cite{gpu-ext} applies eBPF to GPU driver-level resource
management; \sysname targets the collective communication library
layer, where policies govern algorithm and protocol selection
rather than device resources. \sysname is complementary: a~policy
can activate synthesized schedules or tuned configurations based
on runtime conditions.

\paragraph{Alternative extension mechanisms.}
WebAssembly~\cite{wasm} provides sandboxed execution but
lacks eBPF's static safety guarantees (termination, bounded
memory access), and incurs higher overhead from runtime bounds
checks. Out-of-process services add microseconds of IPC
latency, an order of magnitude more than \sysname's
80--130\ns overhead. Declarative DSLs cannot express stateful
policies or feedback loops. Hardware isolation
(Intel MPK~\cite{erim}) provides memory separation without
verification. eBPF uniquely combines static verification with
near-native JIT performance in a single model.

\section{Discussion}
\label{sec:discussion}

\sysname currently covers tuner, profiler, and net plugins;
extending coverage to the env plugin is straightforward. Our
net plugin prototype wraps the built-in Socket backend; deeper
integration (e.g., RDMA-aware policies) is future work. Our
evaluation uses 8~GPUs on a single node with NVLink; multi-node
experiments with InfiniBand and larger rank counts are needed to
validate the approach at production scale. A higher-level policy
DSL compiled to BPF would lower the barrier for ML engineers who
are not familiar with eBPF programming.

The eBPF-for-extensions pattern may generalize beyond NCCL.
AMD's RCCL~\cite{rccl} exposes a similar plugin architecture,
suggesting that a cross-vendor port may be feasible. Other
libraries with native-code extension points (MPI implementations,
RDMA middleware) face similar safety gaps and could benefit from
the same verify-then-execute approach.

\section{Conclusion}
\label{sec:conclusion}

We presented \sysname, a system that transforms NCCL's unsafe
native plugin extension points into verified, composable policy
hooks by embedding a userspace eBPF runtime. NCCL's plugin
architecture is structurally analogous to kernel extension points
where eBPF is already proven; \sysname brings this model to GPU
collective communication, covering tuner, profiler, and net
plugins. \sysname adds 80--130\ns per policy decision and
negligible overhead on the net data path, catches crash-inducing
bugs at load time, supports atomic hot-reload without call loss,
and enables a message-size-aware policy that improves 8-GPU
NVLink AllReduce throughput by up to 27\% over NCCL's default, with
zero NCCL modifications.

\begin{acks}
We thank Verda for providing the 8$\times$~NVIDIA~B300 NVLink
compute infrastructure used in this work.
\end{acks}

\balance
\bibliographystyle{ACM-Reference-Format}
\input{paper.bbl}

\end{document}

%% file: figures/architecture.tex
\begin{tikzpicture}[font=\small, >=Latex,
  every node/.style={inner sep=4pt}]

  \pgfmathsetmacro{\fw}{16.5}

  \draw[nccl soft box, fill=black!1] (0,0) rectangle (\fw cm,4.5cm);
  \node[anchor=north west, font=\small\bfseries] at (0.2cm,4.35cm) {NCCL process};

  \node[nccl box, fill=ncclblue!9, minimum width=2.6cm, minimum height=2.8cm]
    (core) at (1.7cm, 2.1cm) {};
  \node[font=\small\bfseries] at (1.7cm, 2.6cm) {NCCL Core};
  \node[font=\footnotesize, text=ncclslate] at (1.7cm, 2.0cm) {collectives,};
  \node[font=\footnotesize, text=ncclslate] at (1.7cm, 1.6cm) {transports};

  \node[nccl box, fill=ncclblue!9, minimum width=2.5cm, minimum height=1.0cm]
    (tuner) at (5.0cm, 3.1cm) {};
  \node[font=\small\bfseries] at (5.0cm, 3.3cm) {Tuner Adapter};
  \node[font=\footnotesize, text=ncclslate] at (5.0cm, 2.85cm) {\texttt{getCollInfo()}};

  \node[nccl box, fill=ncclteal!12, minimum width=2.5cm, minimum height=1.0cm]
    (prof) at (5.0cm, 1.1cm) {};
  \node[font=\small\bfseries] at (5.0cm, 1.35cm) {Profiler Adapter};
  \node[font=\footnotesize, text=ncclslate] at (5.0cm, 0.9cm) {event callbacks};

  \draw[nccl flow] (core.east |- tuner) -- node[above, font=\footnotesize, text=ncclslate] {metadata} (tuner.west);
  \draw[nccl feedback] (core.east |- prof) -- node[above, font=\footnotesize, text=ncclslate] {timestamps} (prof.west);

  \draw[nccl soft box, fill=ncclblue!4, draw=ncclblue!40]
    (7.0cm, 0.25cm) rectangle (12.0cm, 4.1cm);
  \node[anchor=north west, font=\small\bfseries, text=ncclink] at (7.2cm, 3.95cm)
    {bpftime runtime};
  \node[font=\footnotesize, text=ncclslate] at (9.5cm, 3.5cm)
    {verifier, JIT, maps, helpers};

  \node[nccl chip, fill=ncclgold!15, draw=ncclgold!70!black, minimum width=4.2cm]
    (maps) at (9.5cm, 2.7cm) {\footnotesize\texttt{telemetry / config maps}};

  \node[nccl box, fill=ncclblue!12, minimum width=4.2cm, minimum height=0.65cm]
    (tunpol) at (9.5cm, 1.8cm) {};
  \node[font=\small\bfseries] at (9.5cm, 1.8cm) {Tuner eBPF policy};

  \node[nccl box, fill=ncclteal!12, minimum width=4.2cm, minimum height=0.65cm]
    (profpol) at (9.5cm, 0.85cm) {};
  \node[font=\small\bfseries] at (9.5cm, 0.85cm) {Profiler eBPF policy};

  \draw[nccl flow] (tuner.east) -- (tunpol.west);
  \draw[nccl feedback] (prof.east) -- (profpol.west);

  \draw[nccl flow, dashed] ([yshift=3pt]tunpol.west) -- node[above, font=\footnotesize, text=ncclslate, pos=0.5] {decision} ([yshift=3pt]tuner.east);

  \draw[nccl feedback, shorten >=1pt] ([xshift=-0.5cm]profpol.north) -- node[left, font=\footnotesize, text=ncclslate] {write} ([xshift=-0.5cm]maps.south);
  \draw[nccl flow, shorten >=1pt] ([xshift=0.5cm]maps.south) -- node[right, font=\footnotesize, text=ncclslate] {read} ([xshift=0.5cm]tunpol.north);

  \node[nccl chip, fill=white, minimum width=1.8cm]
    (elf) at (13.8cm, 3.3cm) {\footnotesize\texttt{policy.bpf.o}};
  \draw[nccl flow] (elf.west) -- node[above, font=\footnotesize, text=ncclslate] {load} (12.0cm, 3.3cm);

  \node[nccl chip, fill=ncclgreen!18, draw=ncclgreen!80!black, minimum width=1.8cm]
    (reload) at (13.8cm, 1.8cm) {\footnotesize hot-reload};
  \draw[nccl alert] (reload.west) -- (12.0cm, 1.8cm);

\end{tikzpicture}%